\newcommand{\Spinfour}{\mbox{$\mbox{\em Spin}(4)$}}
\newcommand{\Invzero}[1]{\mbox{\em Inv}({#1})}
\newcommand{\Invtwo}[1]{\mbox{$\mbox{\em Inv}_{SU(2)}({#1})$}}
\newcommand{\Invfour}[1]{\mbox{$\mbox{\em Inv}_{\mbox{\scriptsize{\em Spin}}(4)}
({#1})$}}
\newcommand{\calH}{{\cal H}}
\newcommand{\One}{{\bf 1}}

\newcommand{\bg}{\beta}

\newcommand{\eg}{\epsilon}

\newcommand{\Sg}{\Sigma}

\newcommand{\be}{\begin{equation}}
\newcommand{\ee}{\end{equation}}
\newcommand{\bearr}{\begin{eqnarray}}
\newcommand{\eearr}{\end{eqnarray}}

\newcommand{\Complex}{{\bf C}}

\newcommand{\bfj}{{\bf j}}
\newcommand{\bfJ}{{\bf J}}
\newcommand{\QED}{\rule{1mm}{3mm}}
\newcommand{\threej}[6]{\protect{\scriptsize\mbox{$\left( \begin{array}{ccc}
      #1 & #2 & #3    \\      #4 & #5 & #6
    \end{array}     \right)$}}}
\newcommand{\bincoeff}[2]{\protect{\scriptsize\mbox{$\left( \begin{array}{c}
       #1      \\      #2
      \end{array}     \right)$}}}

\newsavebox{\Xtree}
\newsavebox{\Ytree}
\newsavebox{\Ztree}

\documentstyle[12pt,psfig]{article}
\begin{document}

\title{On relativistic spin network vertices}
\author{Michael P. Reisenberger\\
 Instituto de F\'{\i}sica, Universidad de la Rep\'{u}blica\\
      Igua 4225 esq. Mataojo, C.P. 11400 Montevideo, Uruguay}
\date{September 22, 1998}
\maketitle

\begin{abstract}
Barrett and Crane have proposed a model of simplicial Euclidean quantum gravity
in which a central role is played by a class of \Spinfour\ spin networks 
called ``relativistic spin networks" which satisfy a series of physically 
motivated constraints. Here a proof is presented that demonstrates that the 
intertwiner of a vertex of such a spin network is uniquely determined, up to 
normalization, by the representations on the incident edges and the
constraints. Moreover, the constraints, which were formulated for four valent 
spin networks only, are extended to networks of arbitrary valence, and the 
generalized relativistic spin networks proposed by Yetter are shown to form 
the entire solution set (mod normalization) of the extended constraints. 
Finally, using the extended constraints, the Barrett-Crane model is 
generalized to arbitrary polyhedral complexes (instead of just simplicial 
complexes) representing spacetime. It is explained how this model, like the 
Barret-Crane model can be derived from BF theory by restricting the sum over 
histories to ones in which the left handed and right handed areas of any 
2-surface are equal. It is known that the solutions of classical 
Euclidean GR form a branch of the stationary points of the BF action with 
respect to variations preserving this condition. 
\end{abstract}

\section{Introduction}

The ``Relativistic spin networks'' defined by Barrett and Crane (BC)
are a fundamental ingredient of their proposal for a simpicial model of 
quantum general relativity in four dimensions. In \cite{BC} the space of 
intertwiners that 
relativistic spin networks are allowed to carry is defined implicitly by a 
set of physically motivated constraints. They also exhibit a single solution 
to these constraints. Soon after Barbieri \cite{Barb} gave a partial proof of 
the uniqueness up to normalization of this solution, which relies on some
unproven hypothesies. BC's constraints apply only to 4-valent
spin networks, but their solution to their constraints has been generalized
in a natural way to arbitrary valence by Yetter \cite{Yetter}, and Barrett
\cite{Barrett98} has given a very transparent characterization of this 
extension in the non q deformed case, $q = 1$.

Here a proof (without auxhiliary assumptions) will be given showing 
that the BC solution is the only one, up to normalization, and similarly 
Yetter's generalization of the solution (for $q = 1$) is the unique solution 
up to normalization of a natural generalization of the BC constraints.
In addition a physically motivated extension of the BC model to polyhedral
complexes is outlined. The BC model can be obtained from a sum over histories
quantization of \Spinfour\ BF theory by restricting the histories to ones that
assign equal (suitably defined) left-handed and right-handed areas to any 
2-surface. At the classical continuum level such a constrained BF theory
does reproduce GR. The solutions of GR form a branch of the stationary points 
of the \Spinfour\ BF action with respect to variations that preserve the 
constraint that the left and right handed areas be equal for all 2-surfaces 
\cite{Rei98}. This procedure for obtaining the BC model generalizes 
straighforwardly to complexes of convex 4-polyhedra.

For background information on spin networks see \cite{Yutsis, Penrose,
RSnet, Baeznet, Rei94}. 

\section{Definition of relativistic spin networks}	\label{def}

I will adopt the following definition of relativistic spin networks,
which generalizes BC's definition from 4-valent vertices to vertices of 
arbitrary valence.
\newline

\noindent {\bf Definition 1}:
{\em A relativistic spin network is a \Spinfour\ spin network
such that 
\begin{itemize}
\item[1] On each edge the left handed spin, $j_L$, and the right handed
spin, $j_R$, are equal.
\item[2] In any expansion of an $n$-valent vertex into a sum of trivalent
trees only trivalent trees with $j_L = j_R$ on each of the internal (virtual)
edges appear.
\end{itemize}}

In general the edges of a spin network carry non-trivial irreducible representations
(irreps) of the gauge group, and the vertices carry intertwiners. The intertwiner
for a vertex can be any invariant tensor of the product representation $R$ formed
by the product of the irreps carried by the incoming edges and the duals of the irreps on
the outgoing edges.\footnote{
The dual of a representation $D$ of a group is the 
representation $D^{-1\,T}$ formed by the transposes of the inverses of the 
representation matrices of $D$. If $D$ is unitary its dual is the complex conjugate
representation $D^*$ formed by complex conjugating each matrix element in the 
representation matrices of $D$.} 
The space of invariant tensors of $R$ will be denoted
$\Invzero{R}$. The {\em evaluation} of a spin network is a complex number 
calculated by contracting the intertwiners of the vertices along the edges. An 
intertwiner carries one index for each incident edge. In the evaluation of a 
spin network the pair of indices associated with each edge (one index lives at 
each end) is contracted, leaving ultimately a single $\Complex$ number. 
In the BC model histories of the gravitational field determine relativistic spin networks
on the boundaries of the 4-simplices that form the simplicial spacetime, and the probability
amplitude of each history is the product of the evaluations of these spin networks (times
some simple further factors associated with the lower dimensional simplices).

Relativistic spin networks are spin networks of \Spinfour\ , the covering group of $SO(4)$.
\Spinfour\ is the product of two $SU(2)$s: $\Spinfour = SU(2)\times SU(2)$, where the first
$SU(2)$ factor will be called $SU(2)_L$, the ``left handed" subgroup, and the second $SU(2)_R$,
the ``right handed" subgroup. This factorization extends to the irreps of \Spinfour\ . These
are tensor products of an irrep of $SU(2)_L$ and an irrep of $SU(2)_R$, and their carrying
spaces, i.e the vector spaces on which they act, are the tensor products of
the carrying spaces of the $SU(2)_L$ and $SU(2)_R$ irreps.\footnote{{\em Proof}: 
The representation matrices $D(g_L,g_R)$ of a representation $D$ of 
$SU(2)_L \times SU(2)_R$ may always be written as $D_L(g_L) D_R(g_R)$ 
where $D_L(g_L) = D(g_L,\One)$ and $D_R(g_R) = D(\One,g_R)$. $D_L$ and 
$D_R$ are of course $SU(2)$ representations. The carrying space $\calH$ 
of $D$ can thus be decomposed into $D_L$ invariant subspaces carrying 
irreps of $SU(2)_L$. The fact that $D_L$ and $D_R$ commute, and Shur's 
lemma, then imply that $D_R$ acts on the subspaces of $\calH$ carrying 
the same irrep of $SU(2)_L$, i.e. eigensubspaces of left handed spin. 
The commutation of $D_L$ and $D_R$ further implies that when $D_R$ is 
reduced in these subspaces they decompose into tensor products of $D_L$ 
and $D_R$ invariant subspaces, and the restriction of $D$ to these are 
tensor products of irreps of $SU(2)_L$ and $SU(2)_R$. The orbit of any 
vector in the carrying space of such a product of irreps clearly spans
all of the carrying space, so the products are irreducible
representations of $SU(2)_L\times SU(2)_R$.\QED}

The factorization of the irreps implies that the product $R$ of irreps incident on a spin 
network vertex also factorizes into a left-handed factor, $R_L$, and a right handed factor, 
$R_R$, and hence that the intertwiner space factorizes according to
\be
\Invfour{R} = \Invfour{R_L\otimes R_R} = \Invtwo{R_L} \otimes \Invtwo{R_R}
\ee
into a tensor product of two $SU(2)$ intertwiner spaces.

$SU(2)$ irreps are determined by their spin, modulo the freedom to change basis 
in the carrying space. \Spinfour\ irreps are therefore characterized in the same 
sense by the spins $(j_L, j_R)$ of their left and right handed factors.
To keep the mathematics as concrete as possibloe it is convenient
to fix the bases in the carrying spaces so that the irreps take a standard form  
completely determined by their spins $(j_L,j_R)$.\footnote{
Note that the evaluation of a spin network is invariant under 
changes of basis in the carrying spaces of the irreps on the edges. 
The contractions of the indices of the intertwiners are all between 
vector indices of some irrep and corresponding covector indices, 
i.e. vector indices of the dual of the irrep.}
We shall adopt as our standard $(j_L,j_R)$ irrep
\be
U^{(j_L)}\otimes U^{(j_R)\,*},
\ee
where for each spin $j$, $U^{(j)}$ is a particular, {\em unitary}, spin $j$
$SU(2)$ irrep fixed once and for all by some conventions,\footnote{
For instance we may choose 
\be	\label{U:convention}
U^{(j)\,m}{}_n(g) = \bincoeff{2j}{j+m}^{-\frac{1}{2}}\bincoeff{2j}{j+n}^{-\frac{1}{2}}
\sum_{\Sg M_i = m,\Sg N_i = n} g^{M_1}{}_{N_1}
... g^{M_{2j}}{}_{N_{2j}}\ \ \forall g\in SU(2).
\ee
Here the indices $M_i$ and $N_i$ range over $\{-\frac{1}{2},\frac{1}{2}\}$.}
and $U^{(j)\,*}$ is its dual irrep. This choice can be made because, firstly, the compactness
of $SU(2)$ implies that its irreps preserve a hermitean inner product, and are thus
unitary in orthonormal bases with respect to this inner product, and secondly because
the dual of a spin $j$ $SU(2)$ irrep is also a spin $j$ irrep, so that the \Spinfour\ 
irrep chosen really has right handed spin $j_R$.

Now let's consider a spin network vertex. First let's define some notation.
Once the bases are fixed the product representation $R$ formed by the incident 
irreps is completely determined by the incident spins and whether edges are
incoming or outgoing. This information can be gathered into two vectors, ${\bf j}_L$
and ${\bf j}_R$, which we shall, in a slight abuse of language, refer to as the vectors
of incident left handed and right handed spins. Each entry of ${\bf j}_L$ corresponds
to an incident edge and consists of the left handed spin, $j_L$, if the edge is
incoming and $-j_L$ if the edge is outgoing. The left handed factor of $R$ is then
$R_L = R(\bfj_L)$ where
\be
R(\bfj) = \bigotimes_n U^{(j_n)}.
\ee
($n$ numbers the edges and we have defined $U^{(-j)} \equiv U^{(j)*}$). 
$\bfj_R$ is defined in complete analogy to $\bfj_L$,
so our conventions for the \Spinfour\ irreps on the edges imply that $R = R(\bfj_L)\otimes
R(\bfj_R)^*$. If $\calH(\bfj)$ is the carrying space of the $SU(2)$
representation $R(\bfj)$, and $\Invtwo{\bfj} \equiv \Invtwo{R(\bfj)}$ is its invariant
subspace then the carrying space of $R = R(\bfj_L)\otimes R(\bfj_R)^*$ is 
$\calH(\bfj_L)\otimes\calH(\bfj_R)^*$ and the \Spinfour\ intertwiner space is
$\Invfour{\bfj_L,\bfj_R} = \Invtwo{\bfj_L}\otimes\Invtwo{\bfj_R}^*$.

The inner product preserved by the unitary representation $R(\bfj)$ establishes a
one to one correspondence between vectors of $\calH(\bfj)^*$ and linear functions 
$\calH(\bfj) \rightarrow \Complex$. A tensor 
$\Psi\in\calH(\bfj_L)\otimes\calH(\bfj_R)^*$ can thus be viewed as a linear mapping 
$\Psi:\calH(\bfj_R) \rightarrow \calH(\bfj_L)$. If the tensor $\Psi$ is an intertwiner 
it maps $\Invtwo{\bfj_R}$ into $\Invtwo{\bfj_L}$ and the orthogonal complement of 
$\Invtwo{\bfj_R}$ in $\calH(\bfj_R)$ to zero. 

Condition 1 in the definition of relativistic spin networks implies that 
$\bfj_L = \bfj_R$ at their vertices. Thus a relativistic
intertwiner $\Phi$ may be viewed as a mapping of $\calH(\bfj_R)$ into {\em itself},
that furthermore maps \Invtwo{\bfj_R}\ into itself and the orthogonal complement of
\Invtwo{\bfj_R}\ to zero. It is the composition of the orthogonal projector
$P$ onto \Invtwo{\bfj_R}\ and a linear mapping $X$ of \Invtwo{\bfj_R}\ to itself: $\Phi = XP$.

Condition 2 in the definition of relativistic spin networks refers to 
trivalent tree expansions of spin network vertices. It is well known 
(see \cite{Yutsis},\cite{RdP}) that for $SU(2)$ spin networks each 
trivalent tree graph having the same external edges as a given vertex 
defines a basis of the intertwiner space \Invtwo{\bfj}\ of the 
vertex.\footnote{Trivalent tree bases exist for spin networks of any group 
for which the product of two irreps is completely reducible.} 
(The trivalent tree has oriented edges and a cyclic ordering of
the edges incident at each vertex.) 
Each element of such a ``trivalent tree basis" is associated with an 
assignment $\bf J$ of (possibly zero) spins to the internal edges
of the tree (also known as ``virtual" edges because they are not present in the actual 
spin network). The basis element is evaluated by contracting the intertwiners 
of the trivalent vertices along the internal edges as in a spin network evaluation. This
leaves just the intertwiner indices associated with the external edges free.
To complete the definition one needs to specify the trivalent intertwiners. The
trivalent intertwiner spaces \Invtwo{j_1,j_2,j_3}\ of $SU(2)$ are all one dimensional,
so it is sufficient to fix the freedom to multiply the intertwiners by scalar 
factors. We will choose the trivalent intertwiners to be normalized. 
Then, if a normalizing factor $\sqrt{2J + 1}$ is included for each internal edge,
the trivalent tree intertwiners will be normalized.
There remains a phase which must be chosen by convention. We will 
suppose that such a convention has been adopted,\footnote{
The Wigner 3-jm symbols, $\threej{j_1}{j_2}{j_3}{m_1}{m_2}{m_3}$, 
with the standard convention \cite{Yutsis} that 
$(-1)^{j_1 - j_2 + j_3}\threej{j_1}{j_2}{j_3}{j_1}{(j_3 - j_1)}{-j_3}$ 
is real and non-negative for all $j_1$, $j_2$, and $j_3$, form a basis
of trivalent intertwiners consistent with the convention for $U^{(j)}$ 
of (\ref{U:convention}). Aside from being real this basis also has the attractive
feature that the intertwiners have simple symmetry properties under permutations
of the incident edges: If $j_1 + j_2 + j_3$ is even they are symmetric, if it is
odd they are antisymmetric.}
so that the trivalent tree $T$ and the vector $\bfJ$ of spins on the 
internal edges determine a unique intertwiner $|T,\bfj,\bfJ\rangle \in\Invtwo{\bfj}$.
\Spinfour\ trivalent tree bases can be then be constructed from the $SU(2)$ trivalent
tree bases: If the trivalent tree $T$ spans a \Spinfour\ vertex with incident spins 
$(\bfj_L,\bfj_R)$ then the multiplet
\be
\{|T,\bfj_L,\bfJ_L\rangle \otimes \langle T,\bfj_R,\bfJ_R|\}_{\bfJ_L,\bfJ_R}
\ee
spans the intertwiner space \Invfour{\bfj_L,\bfj_R}\ . Here $\langle T,\bfj,\bfJ|$
denotes the complex conjugate of the tensor $|T,\bfj,\bfJ\rangle$, which, as has 
been explained, defines a linear function $\calH(\bfj) \rightarrow \Complex$ via the 
hermitean inner product, thus justifying the Dirac bra notation.
The definition of a relativistic spin network implies that the expansion 
of an intertwiner $\Phi$ of such a spin network on a trivalent tree basis has the form
\be		\label{tritreeexp}
\Phi = \sum_{\bfJ} a^T_{\bfj\,\bfJ}\: |T,\bfj,\bfJ\rangle \otimes \langle T,\bfj,\bfJ|,
\ee
i.e. the left and right handed spins are equal on both external and internal edges, for
{\em any} tree $T$ spanning the vertex. 

\section{The solution to the constraints defining relativstic vertices}

Now that relativistic spin networks have been defined and explained
we are ready to state and prove our result:
\newline

\noindent {\bf Theorem}
{\em The intertwiner of any vertex in a relativistic spin network is uniquely
determined, up to a numerical factor, by the irreps on the incident edges. 
Let $\bfj = \bfj_L = \bfj_R$ be the common value of the vectors of left and 
right handed incident spins (as defined in \S \ref{def}) at a vertex, and let 
$\calH(\bfj)$ be the space of tensors transforming under the product of the 
$SU(2)$ irreps with these spins, equipped with the Hermitean inner product 
preserved by $SU(2)$. When the bases in the carrying spaces of the $SU(2)_L$
and $SU(2)_R$ irreps on the edges of the relativistic spin network are chosen 
so that the $SU(2)_L$ irrep is the dual of the $SU(2)_R$ irrep on each edge 
incident on the vertex then
\begin{itemize}
	\item[1] tensors, like the intertwiner, that transform under the 
product of the incident \Spinfour\ irreps live in the tensor product of 
$\calH(\bfj)$ and its dual, and can thus be viewed as linear mappings of 
$\calH(\bfj)$ to itself, and
	\item[2] the intertwiner is proportional to the orthogonal projector 
from $\calH(\bfj)$ to the subspace of invariant tensors $\Invtwo{\bfj} 
\subset \calH(\bfj)$.
\end{itemize}}

\noindent {\em Proof}:
Part 1 has already been established in \S \ref{def}, so only part 2 remains 
to be proven. (\ref{tritreeexp}) shows that an intertwiner, $\Phi$, of 
a relativistic spin network is the composition of the projector $P$ from the 
space $\calH(\bfj)$ of tensors with spins $\bfj$ onto the subspace of 
intertwiners $\Invtwo{\bfj}$ and a linear map $X$ of $\Invtwo{\bfj}$ to 
itself. (\ref{tritreeexp}) furthermore requires that $X$ is {\em diagonal} in 
each of the trivalent tree bases of $\Invtwo{\bfj}$. Obviously 
$X = c\One$ with $c \in \Complex$ satisfies this condition,
so $c P$ is a relativistic intertwiner. To establish the theorem it
remains to be shown that the set of all trivalent tree bases
is rich enough so that this is the only solution to the condition.

\sbox{\Xtree}
{\begin{picture}(14,9)
\put(0,0){\line(1,1){4}}
\put(4,4){\line(-1,1){4}}
\put(4,4){\line(1,0){6}}
\put(10,4){\line(1,1){4}}
\put(10,4){\line(1,-1){4}}
\end{picture}}
\sbox{\Ytree}
{\begin{picture}(9,14)
\put(0,0){\line(1,1){4}}
\put(4,4){\line(1,-1){4}}
\put(4,4){\line(0,1){6}}
\put(4,10){\line(1,1){4}}
\put(4,10){\line(-1,1){4}}
\end{picture}}
\sbox{\Ztree}
{\begin{picture}(10,10)
\put(0,0){\line(1,1){10}}
\put(3,3){\line(1,0){4}}
\put(10,0){\line(-1,1){4}}
\put(4,6){\line(-1,1){4}}
\end{picture}}
\newcommand{\Xintw}[1]
{\begin{picture}(14,9)
\put(0,-3){\usebox{\Xtree}}
\put(6,2){#1}
\end{picture}}
\newcommand{\Yintw}[1]
{\begin{picture}(9,14)
\put(0,-6){\usebox{\Ytree}}
\put(5,0){#1}
\end{picture}}

Let us first consider four valent vertices. There are three (unoriented) 
trivalent trees matching the four incident edges, 
\begin{center}
\raisebox{-6mm}{\usebox{\Ytree}}, \raisebox{-4mm}{\usebox{\Ztree}}, 
and \raisebox{-3mm}{\usebox{\Xtree}}
\end{center}
each of which has one internal, or virtual, edge. Once the orientations of 
the external edges are fixed to match those of the four valent vertex being 
expanded there remains the freedom to choose the orientation of the internal 
edge and the cyclic ordering of the incident edges at the two trivalent 
vertices. However these choices only affect the {\em signs} of the 
corresponding trivalent tree basis, modulo sign it is determined by the 
unoriented tree graph.

Let's number the incident edges 1,2,3,4 clockwise from the top left, and let 
$G_{n\,i}$ for $n\in\{1,2,3,4\}$ be the generators,\footnote{
If the edge $n$ is incoming then $[G_{n\,i},G_{n\,j}] = i \eg_{ij}{}^k 
G_{n\,k}$. If the edge is outgoing the negatives $-G_{n\,i}$ satisfy these 
commutation relations. The generators belonging to distinct edges of course 
commute.}
acting in $\calH(\bfj)$, of the $SU(2)$ irreps on the incident edges. Let's 
also define $G_{mn\,i} = (G_m + G_n)_i$. The first of the trivalent trees 
drawn above corresponds to a pairing of edges 1 and 2, which join at a 
trivalent vertex. The corresponding basis intertwiners 
\rule[-6mm]{0mm}{14mm}\Yintw{$J$} is a contraction
of a trivalent intertwiner at this vertex and one at the other vertex of the 
graph. The invariance of the intertwiner at the first vertex implies that
\be
G_{12}^2 \rule[-6mm]{0mm}{14mm}\Yintw{$J$} = J(J+1)\rule[-6mm]{0mm}{14mm}
\Yintw{$J$}.
\ee
The intertwiner basis $\{\rule[-6mm]{0mm}{14mm}\Yintw{$J$}\}_J$ is thus the 
eigenbasis of $G_{12}^2$ in $\Invtwo{\bfj}$. Similarly the trivalent tree 
bases associated with the second and third trees diagonalize $G_{13}^2$ and 
$G_{14}^2$ respectively.

Since $X$, the restriction of $\Phi$ to $\Invtwo{\bfj}$, is diagonal in all 
the trivalent tree bases it commutes with $G_{12}^2$, $G_{13}^2$, and 
$G_{14}^2$. What's more, since the spectra of these operators 
($\{J(J+1)\}$) are non-degenerate, $X$ can be expressed as a function of any 
one of them. Choosing $G_{12}^2$ we write
\be
X = \sum_{q = 0}^{d-1} b_q [G_{12}^2]^q,
\ee
where $d$ is the dimension of $\Invtwo{\bfj}$. Since the spin $J$ defined by 
the eigenvalues $J(J+1)$ of $G_{12}^2$ can take only $d$ values a polynomial 
of degree $d-1$ can reproduce any dependence of $X$ on $G_{12}^2$.

Because $X$ also commutes with $G_{12}^2$
\be		\label{comm}
0  = [G_{13}^2,X] = \sum_{q=0}^{d-1} \:b_q\: [G_{13}^2,(G_{12}^2)^q]
\ee
This condition implies that $b_q = 0 \ \forall q\neq 0$, so that 
$X = b_0 \One$. To prove this it is sufficient to consider the matrix 
elements of (\ref{comm}) between the basis intertwiners
\rule[-6mm]{0mm}{14mm}\Yintw{$J$} and 
\rule[-6mm]{0mm}{14mm}\Yintw{$J-1$}\ \ \ \ (which I will denote 
$|J\rangle$ and $|J-1\rangle$ in the following) for all 
allowed values of $J$. $\langle J|[G_{13}^2,(G_{12}^2)^q]|J-1\rangle$ is 
obviously zero when $q = 0$. When $q\geq 1$ it equals
\be
\sum_{r = 0}^{q-1}\:\langle J|(G_{12}^2)^r\,[G_{13}^2,G_{12}^2]\,
(G_{12}^2)^{q-1-r}|J-1\rangle = \bg_J P_q(J),
\ee
where $\bg_J = \langle J|[G_{13}^2,G_{12}^2]|J-1\rangle$, and 
\be
P_q(J) = \sum_{r = 0}^{q-1}\:[(J)(J+1)]^r [(J-1)J]^{q-1-r}.
\ee
(\ref{comm}) therefore implies 
\be		\label{constr}
0 = \bg_J \:\sum^{d-1}_{q = 1} b_q P_q(J).
\ee

The matrix elements of $[G_{13}^2,G_{12}^2]$ have been worked out explicitly 
by Lev\'y-Leblond and Lev\'y-Nahas \cite{LL65}.\footnote{
$[G_{13}^2,G_{12}^2] = -4i\eg^{ijk}G_{1\,i}G_{2\,j}G_{3\,k}$. It is thus $-4i$
times the squared 3-volume operator associated with the four valent vertex in 
loop quantum gravity \cite{discarea2,AL96b,Barbvol}.}
They find\footnote{
(\ref{matrixel}) corresponds to equation (2.17) of \cite{LL65}. Their $J$ is 
called $j_4$ in our notation and their $l$ is our $J$. Our formula has
an extra factor $(2j_4 + 1)$ relative to theirs because, while they are
evaluating the matrix element between states of {\em three} spins having 
definite values ($j_4$ and $m_4$) of the magnitude and $3$ axis component of
the total spin, we are calculating the matrix elements between states
of {\em four} spins which have total spin zero. In this latter calculation
one must sum over the $2j_4 + 1$ possible values of $m_4$.}
{\samepage
\bearr		\label{matrixel}
\bg_J = \frac{2j_4 + 1}{\sqrt{4 J^2 - 1}} &\!\!\!&
\left\{[(j_1 + j_2 + 1)^2 - J^2][J^2 - (j_1 - j_2)^2]\right\}^{\frac{1}{2}} 
\nonumber \\
&& \times \left\{[(j_3 + j_4 + 1)^2 - J^2][J^2 - (j_3 - j_4)^2]\right\}^
{\frac{1}{2}}
\eearr}

The important feature of this expression for us is that it is non-zero for 
all values of $J$ such that both $J$ and $J-1$ satisfy the triangle 
inequalities 
\bearr
|j_1 - j_2| \leq & \mbox{spin} & \leq j_1 + j_2	\\
|j_3 - j_4| \leq & \mbox{spin} & \leq j_3 + j_4
\eearr
for the spin on the internal edge of \raisebox{-6mm}{\usebox{\Ytree}}. 
It is thus non-zero for all but the smallest of the values
of $J$ corresponding to the intertwiners $|J\rangle$ spanning $\Invtwo{\bfj}$.
The condition (\ref{constr}) therefore implies
\be		\label{constraint}
0 = \sum^{d-1}_{q = 1} b_q P_q(J)
\ee
for all but one of the $d$ allowed values of $J$. 

Now note that, firstly, the highest order term in $P_q(J)$ is $J^{2(q-1)}$, 
and, secondly, that $P_q(-J) = P_q(J)$. It follows that the $P_q$ have the 
form $P_q(J) = \sum_{p = 0}^{d-2} A_{qp} J^{2p}$, where $A$ is a 
$(d-1)\times (d-1)$ matrix. Moreover, since their leading powers of $J$ are
all distinct, the $P_q$ are linearly independent polynomials, implying that 
$A$ is invertible. From (\ref{constraint}) it follows that 
$\sum_{q=1}^{d-1} b_q A_{qp} = 0\ \forall p\in\{0,...,d-2\}$. The 
invertibility of $A$ then shows that $b_q = 0\ \forall q\in \{1,...,d-1\}$. 
As claimed $X = b_0\One$.

\begin{figure}
\centerline{\psfig{figure=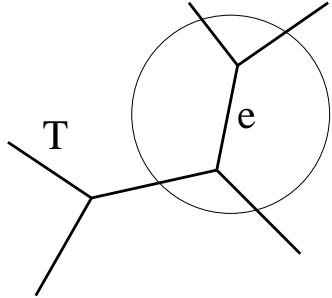,height=4cm}}
\caption[xxx]{The circle contains a four valent fragment of the trivalent
tree graph $T$}
\label{tree}
\end{figure}

The result is thus established for four valent vertices. Let's now consider
a vertex of arbitrary valence. Fix a particular trivalent tree graph $T$
expanding the vertex and consider a four valent fragment of the
graph consisting of an internal edge $e$, and the four (internal or external)
edges attached to it (see Figure \ref{tree}). The preceeding arguments
relating to four valent vertices can be applied directly to this fragment and 
show that the eigenvalue $a^T_{\bfj\,\bfJ}$ of X on $|T,\bfj,{\bf J}\rangle$ 
is independent of the spin on $e$. Since this holds for any internal edge of 
$T$ the eigenvalue is independent of all the internal spins, i.e it has a 
common value $c \in \Complex$ on each element of the basis 
$\{|T,\bfj,{\bf J}\rangle\}_{\bfJ}$. Hence $X = c\One$ and the relativistic 
intertwiner is $cP$.\QED
\newline

I close with a few observations 
\begin{itemize}
\item The BC model has a simple, physically motivated extension to arbitrary 
polyhedral complexes (as opposed to simplicial complexes) representing spacetime.
The generalized BC constraints of definition 1 are equivalent to the requirement
that relativistic spin networks define equal left and right handed areas
for any surface, including ones cutting through vertices. Here the left and
right handed areas are determined from the left and right handed spins on the
spin network edges and on the virtual edges of trivalent tree expansions of the
vertices using the the area operator of loop quantum gravity \cite{discarea2, AL96a, Fritelli96}.

In \cite{Rei98} it has been shown that GR is a branch of the theory 
obtained by restricting $SO(4)$ (or \Spinfour\ ) BF theory to histories in 
which left handed and right handed areas are equal. That is, the solutions of 
GR form a branch of the stationary points of the $SO(4)$ BF action with respect 
to variations that respect the constraint that left handed areas equal right handed areas. 
This provides a motivation of the BC model which may be extended to polyhedral complexes:
\Spinfour\ BF theory is just two non-interacting $SU(2)$ BF theories, corresponding to
$SU(2)_L$ and $SU(2)_R$ respectively. Thus Ooguri's \cite{Ooguri} simplicial lattice 
sum over histories quantization of $SU(2)$ BF theory\footnote{
This model is also known as the Crane-Kauffman-Yetter model because Crane, 
Kauffman, and Yetter \cite{CKY} regulated it by q deforming it and were then able to 
show that the resulting regulated model really
defines a continuum (lattice independent) topological field theory.}
immediately provides a simplicial 
quantization of \Spinfour\ BF theory. The BC model is then obtained from simplicial 
\Spinfour\ BF theory by restricting the histories to ones in which left and right handed 
areas are equal. A history in the \Spinfour\ BF theory defines a \Spinfour\ spin network
on the boundary of each 4-simplex (or more precisely, on the 1-skeleton of the dual
of the boundary seen as a three dimensional simplicial complex), which plays the role of 
boundary data in the sense that the spins and intertwiners of the spin networks on two 
neighboring 4-simplices must match in their mutual boundary. The requirement that left and 
right handed areas be equal then reduces the allowed \Spinfour\ spin networks to just 
relativistic spin networks. 

Ooguri's quantization of BF theory is most easily generalized to
arbitrary polyhedral complexes in the connection formulation (see \cite{Rei97} for a detailed
discussion). In this formulation 
the boundary data on each 4-cell is a lattice connection (of \Spinfour\ in our case) 
defined by the parallel transport matrices across the 2-cells separating the 3-cells
of the boundary of the 4-cell. (Equivalently, it is a lattice connection on the 1-skeleton
of the dual of the boundary). The amplitude of this connection is a delta distribution with
support on flat connections. Clearly the sum over histories yields the same states
on the boundary of the spacetime (once infinities stemming from redundancies in the delta 
functions are factored out) whether simplices or arbitrary convex polyhedral 4-cells
form the spacetime complex. 

If one transforms the sum over connection boundary data
to a sum over spin network boundary data (see \cite{Rei97}) one finds, for polyhedral
complexes as for simplicial complexes, that the amplitudes of each history is just
the product of the evaluations of the spin networks on the 4-cells, times a factor
$(2j + 1)^2$ for each 2-cell, with $j = j_L = j_R$ the common value of the spins
carried by the spin network edges crossing that 2-cell. Applying the constraint
that left areas equal right areas for all surfaces, even ones crossing the vertices
of the spin networks, restricts the spin networks on the boundaries of the 4-cells
to be relativistic spin networks in the extended sense of our definition 1. This
defines the generalization of the BC model to polyhedral complexes.

\item To completely determine the BC sum over histories a normalization has to
be chosen for the relativitic intertwiners. The four valent relativistic intertwiner 
given by BC in \cite{BC} seems to be just $P$. On the other hand, if the sum over histories
is to be truly a restriction of the sum over histories for BF theory then the relativistic
intertwiner must be normalized in the sense that its contraction on all indices with
its complex conjugate must be $1$. Thus it must be $P/\sqrt{d}$ up to a phase,
where $d$ is the dimensionality of $\Invtwo{\bfj}$. 

\item Barrett \cite{Barrett98} has shown that Yetter's extension to
arbitrary valence of the four valent relativistic intertwiner found
by BC is equal to 
\be		\label{intformula}
\int_{SU(2)} dg \prod_{j \in {\bf j}} U^{(j)}(g)
\ee
(when the conventions fixing the bases in the irrep carrying spaces are
adopted). Here $U^{(j)}(g)$ is the spin $j$ representation matrix of $g \in SU(2)$
corresponding to the basis convention, and the normalized Haar measure 
is used to integrate over the group.
(\ref{intformula}) is precisely the orthogonal projector $P$ on $\Invtwo{\bf j}$. 
Thus the theorem shows that the unique solution (mod normalization) of 
our generalization of BC's constraints is Yetter's extension of their
solution.

\item Since $j_R$ is everywhere equal to $j_L$ the BC state sum can be viewed
as a sum over histories of left handed ``fields'', i.e. the $j_L$ only. Thus the BC model
can be viewed, like the model of \cite{Rei97} as a (proposal for) a formulation of 
quantum GR in terms of ``self-dual" variables. (Note that this does {\em not} mean that
it is a model of only the self-dual sector of GR, in which the anti-self dual curvature 
vanishes. It only means that exclusively self-daul variables are used to express the
configuration of the gravitational field.)   
\end{itemize}
 
\section*{Acknowledgements}

I would like to thank John Barrett, Rodolfo Gambini, and Carlo Rovelli for encouragement 
and discussions regarding the subject of this note and closely related issues. This
work was funded by the Comision Sectorial de Investigaci\'on Cientifica (CSIC) of the
Universidad de la Rep\'ublica.

\end{document}